\documentclass[12pt]{article}
\input epsf.sty
\usepackage{pdfpages}
\newcommand{\half}{{1\over 2}}

\newcommand{\nl}{\hspace{-.65cm}}
\newcommand{\be}{\begin{equation}}
\newcommand{\ee}{\end{equation}}
\newcommand{\ben}{\begin{eqnarray}\displaystyle}
\newcommand{\een}{\end{eqnarray}}

\newcommand{\sectiono}[1]{\section{#1}\setcounter{equation}{0}}

\def\sqr#1#2{{\vcenter{\vbox{\hrule height.#2pt
         \hbox{\vrule width.#2pt height#1pt \kern#1pt
            \vrule width.#2pt}
         \hrule height.#2pt}}}}

\begin{document}

\begin{center}
\large{\bf  STRING THEORY AT THE TIP OF THE CIGAR}

\vspace{10mm}

\normalsize{Amit Giveon \\\vspace{2mm}{\em Racah Institute of Physics, The Hebrew University, Jerusalem, 91904, Israel} \\\vspace{2mm}{\em and} \\\vspace{2mm}  Nissan Itzhaki \\\vspace{2mm}{\em Physics Department, Tel-Aviv University,
Ramat-Aviv, 69978, Israel}}


\end{center}

\vspace{10mm}

\begin{abstract}

In arXiv:1208.3930
we argued that string theory at the tip of the cigar geometry, obtained by a Wick rotation of the Schwarzschild background, exhibits unexpected features that include degrees of freedom that are confined to the tip.
Using the exact worldsheet description of near extremal NS5-branes
we further study string theory at the tip of the cigar and find more evidence to its surprising properties.

\end{abstract}

\newpage

\baselineskip=18pt


\newpage
\renewcommand{\theequation}{\thesection.\arabic{equation}}
\sectiono{Introduction }
\bigskip
Recently \cite{Giveon:2012kp},
we argued that the tip of the cigar geometry, obtained by Wick rotation of a large Schwarzschild black hole, is special in {\em classical} string theory. The sense in which it is special is that there are stringy degrees of freedom, that are absent in General Relativity, which are confined to the tip.  In the absence of an exact worldsheet description of the Schwarzschild geometry, to reach this radical conclusion, we had to rely on a certain effective action,
whose validity and interpretation are questionable.

For the following two reasons,
the near horizon geometry associated with $k$ near extremal NS5-branes is an ideal system to test the conclusions of \cite{Giveon:2012kp}.
First, this system  has an exact worldsheet description.
Second, like Schwarzschild black holes
and unlike large black holes in AdS geometry,  the Hawking process will radiate the energy above extremality to infinity  \cite{Maldacena:1997cg}.
Therefore, if something unexpected happens at the tip associated with a Schwarzschild black hole
then it should also be found at the tip of the cigar associated with NS5-branes.
Our goal here is to take advantage of the exact worldsheet description to explore the stringy physics at the tip of the cigar of NS5-branes.

The paper is organized as follows.
In section 2, we recall the reasoning of \cite{Giveon:2012kp}.
Section 3 contains a short review of the relevant facts concerning
the exact worldsheet description of the near horizon geometry associated with $k$ near extremal NS5-branes.
In sections 4 and 5, we use  this exact worldsheet description to study the properties of string theory at the tip of the cigar.
We conclude in section 6, and some details are presented in the appendix.

\sectiono{The basic idea }
\bigskip
In this section, we recall the basic idea of \cite{Giveon:2012kp}
that argued that the tip of the cigar is special in string theory.

Consider string theory on the Euclidean version of the black hole -- the cigar background.
For concreteness, we focus on the cigar background  obtained by Wick rotating the Schwarzschild solution in four dimensions,
\be\label{bg}
ds^2=\left(1-\frac{2GM}{r}\right)dx_0^2+\frac{dr^2}{1-\frac{2GM}{r}}+r^2d\Omega_2^2.
\ee
The periodicity $x_0\sim x_0 +8\pi G M$ ensures that there is no singularity at the tip, $r=2GM$.
As a result, in general relativity the tip is not special -- the curvature there scales like $1/M^2$,
and   is small for a large black hole.

Globally, the tip is special since it is the only  point invariant under translation in $x^0$.
The argument why string theory is sensitive to this global aspect of the solution is the following.
Far from the tip, (\ref{bg})  describes a semi-infinite cylinder. The mutual locality condition is
\be\label{locality}
(F_1 \alpha_2 - F_2 \alpha_1) - (\bar{F}_1 \bar{\alpha}_2 - \bar{F}_2 \bar{\alpha}_1)+2(p_1 w_2+p_2 w_1)\in 2 Z,
\ee
where $p$ and $w$ denote the momentum and winding numbers and we are using the notation of \cite{polchinskibook}:  $F$ is the worldsheet fermion number and $\alpha= 0 ~(1)$ in the NS (R) sector.

Since $x^0$ is a thermal direction   $p\in Z + \frac12(\alpha-\bar\alpha)$  and  the mutual locality condition (\ref{locality}) reads
\be
\left((F_1+w_1) \alpha_2 - (F_2+w_2) \alpha_1\right) - \left((\bar{F}_1+w_1) \bar{\alpha}_2 - (\bar{F}_2+w_2) \bar{\alpha}_1\right)\in 2 Z.
\ee
Hence, the GSO projection depends on $w$: in, say, type IIB it is \cite{Atick:1988si}
\be
(-)^{F+w}=(-)^{\bar{F}+w}=1.
\ee
In particular, this means  that the spectrum does include tachyons as long as they have odd winding number.
Due to the winding, the mass of these tachyonic modes is large far from the tip and they can be viewed as a  non-perturbative $\alpha'$ correction that roughly scales like $\exp(-M)$.

However, as we approach the tip they become lighter and could potentially affect the low energy physics.
To test this, we follow \cite{Kutasov:2005rr}
and  write down an effective action that describes these modes. We work in the large $M$ limit and focus on the near tip region,
\be\label{or}
ds^2=d\rho^2+\rho^2 d\phi^2,
\ee
where $\phi=x^0/4GM$  has the usual periodicity $\phi\sim\phi+2\pi$, and $\rho$ is the invariant distance from the tip of the cigar.
In units where $\alpha'=2$, the effective action that describes a tachyon with winding number $w$ is
\be\label{ac}
S=\frac12 \int_{0}^{\infty}d\rho \rho\left( (\partial_{\rho} T)^2 +m^2(\rho) T^2 \right), ~~~w\in 2Z+1~,
\ee
with
\be\label{ca}
m^2(\rho)=(-1+w^2\rho^2/4).
\ee
Eqs. (\ref{ac}, \ref{ca}) describe a 2D harmonic oscillator with the following energy levels,
\be\label{har}
E(w,n_1,n_2)= -1 + (n_1+n_2+1)|w|, ~~~n_1,n_2=0,1,2...~.
\ee
At level $n=n_1+n_2$, the $n+1$ degeneracy is resolved by the angular momentum
\be\label{o}
L=n_1-n_2=n,n-2,...,2-n,-n.
\ee

The action (\ref{ac}) is obtained by working in the gauge $\rho=\tau_2$ and $\phi=\tau_1$.
Thus, the target space quantities, $E$ and $L$, have a simple worldsheet interpretation. In particular, the left and right dimensions are
\be\label{r}
(h,\bar{h})=((n_1 +1/2) |w| ,~ (n_2 +1/2 )|w|).
\ee
These conformal dimensions are the same as the ones in type II strings on flat space. However, these modes are localized at the tip.
Namely, the situation appears to be similar to that of an orbifold:  on top of the modes that can propagate everywhere in space there are modes that are confined to the tip.

Interestingly,
for $w=\pm 1$ the ground state of the 2D harmonic oscillator has zero energy and angular momentum. From the worldsheet point of view this implies a  zero-mode, $(h,\bar{h})=(1/2,1/2)$.
Typically, in superstrings on flat space-time  the only zero modes are space-filling.
Here, however, because of the unusual GSO projection,
we encounter a {\em localized} zero-mode. The zero-mode wave function is
\be\label{ta}
T(w=\pm 1, n_1=0,n_2=0)\sim\exp(-\rho^2/4)~,
\ee
which implies that the localization scale is the string scale.

\subsection{ Is this right?}

There are reasons to doubt the argument presented above. In particular, the validity and interpretation of the effective action  (\ref{ac}) are questionable.
Roughly speaking, one can divide the arguments against (\ref{ac}) into three categories:\footnote{We thank D. Kutasov and J. Maldacena for discussions on these issues.}
 \\
1. The effective action uses the winding number, $w$, to define sectors in the theory. This makes perfect sense at the semi-infinite cylinder, where $w$ is conserved. But near the tip, where the zero mode lives, $w$ is not conserved and  it is not clear if a $w$ dependent action is meaningful.\\
2. Should we trust the exact form of  (\ref{ac})? This form is the most naive form in which we simply added the classical contribution $(2\pi \rho T)^2 $ to the ground state energy $-2/\alpha'$. One can easily imagine corrections to this action that will change our conclusion  that there is a localized zero-mode. \\
3. Even if (\ref{ac}) is technically exact,
is it clear that this does not happen around any point in $R^2$?
Namely, could it be that the zero-modes we discuss are related to the usual zero modes associated with a constant dilaton and B-field by averaging over $R^2$?

Luckily, there is a setup in which we can address these issues in a rather precise way. The setup is the near horizon limit of $k$ near extremal NS5-branes. The advantage of considering this system is that it has an exact CFT description which  allows us to address the issues above. Moreover, unlike large black holes in AdS spaces, near extremal NS5-branes  radiate away all of the energy above extremality   in the near horizon limit~\cite{Maldacena:1997cg}. The reason is that, like in Minkowski  and unlike in $AdS$, the time it takes the radiation to reach the boundary is infinite. Hence, if the tip of  a Schwarzschild black hole is special then it should be special also here.

\sectiono{Generalities of $SL(2)/U(1)$ }
\bigskip
The near horizon Euclidean
geometry associated with $k$ near extremal NS5-branes
in the type II superstring is (we work with $\alpha'=2$) \cite{Maldacena:1997cg}
\ben\label{cigar}
ds^2&=&\tanh^2 \left(\frac{\rho}{\sqrt{2 k}}\right) dx_0^2+d\rho^2 +2 k d\Omega_3^2 +(dy_1^2+...dy_5^2),\\ \nonumber
\exp(2 \Phi )&=& g_{0}^2\frac{1}{\cosh^2 \left(\frac{\rho}{\sqrt{2 k}}\right)},~~~~~g_{0}^2=\frac{k}{\mu},
\een
where $\mu$ is the energy density above extremality, and $g_0$ is the value of the string coupling at the
tip of the two-dimensional cigar geometry.
Much like in the Schwarzschild case, our main interest here is in the 2D geometry (spanned by $x_0$ and $\rho$)
that was found long ago \cite{Elitzur:1991cb,Mandal:1991tz,Witten:1991yr,Dijkgraaf:1991ba}.
This solution is perturbatively exact in $\alpha'$~\cite{Bars:1992sr,Tseytlin:1993my}.
We refer to the region $0\leq\rho ~\tilde{<} ~\sqrt{k}$ as the cup of the cigar
and $0\leq\rho ~\tilde{<} ~1$ as the tip of the cigar.

Smoothness of the background at the tip implies that the asymptotic radius of the cigar is $\sqrt{2 k}$.
Thus, the Hawking Temperature is $(2\pi\sqrt{2 k})^{-1}$.
Since the periodicity of $x_0$ does not depend on $\mu$, there is a zero mode associated with changing $\mu$
\be\label{ze}
 \mu\partial_{\mu} ~~\Rightarrow ~~  -\exp(2 \Phi ).
\ee
For any finite $k$ this mode is normalizable.

We are interested in taking $k\gg1 $ and $\mu/k \gg 1$. In this case,
the curvature and string coupling constant are small everywhere and, much like in the case of (\ref{bg}) with large $M$, the tip of the cigar is not special at the SUGRA level.
Taking $k\to \infty$,
the cup of the cigar becomes $R^2$, and (\ref{ze}) becomes the delta-normalizable mode associated with a constant shift of the dilaton.

The question is whether something special happens at the tip in string theory?  In particular, are there localized modes that are confined to the origin as is argued in section 2?
The advantage of this setup over the one of section 2 is that
string theory on this background is described by an exact CFT,
\be\label{slsut}
SL(2,R)_{k}/U(1) \times SU(2)_{k} \times T^5,
\ee
and so this question can be addressed at a more rigourous level. Below we review some of the properties of this model that are relevant for our discussion;
more details can be found e.g. in~\cite{Giveon:2003wn,Aharony:2004xn} and references therein.

Vertex operators in the $SL(2,R)/U(1)$ $N=2$ SCFT are determined by
five quantum numbers ($n_f, \bar{n}_f, m, \bar{m}, j$);
$n_f$ ($\bar{n}_f$) are related to $F$ and $\alpha$ defined in the previous section in the following way
\be
n_f=F+ {1\over 2}\alpha -1 ,~~~~ \bar{n}_f=\bar{F}+\half\bar{\alpha} -1,
\ee
and $m, \bar{m}$ are related to the momentum $p$ and winding $w$ around $x_0$ via
(see e.g. \cite{Giveon:2003wn} for details)
\be
(m+n_f, \bar{m}+\bar{n}_f)=\frac12 (p+kw,-p+kw),~\qquad p\in Z+n_f-\bar n_f ,w\in Z.
\ee
Roughly speaking, $j$ is associated with the momentum in the radial direction $\rho$
of the cigar.

More precisely, the possible values of $j$ are inherited from either
the continuous or the discrete representations of $SL(2)$.
In the continuous representations,
\be\label{js}
j=-1/2+is, ~~s\in R,
\ee
while in the discrete representations,~\footnote{Here, $m>0$, w.l.g.}
\be
j=m-n, ~~~~n=1,2,...~.
\ee
Quantum mechanically, we have to impose the unitarity bound \cite{Giveon:1999px,Maldacena:2000hw}
\be
-\frac12 < j < \frac{k-1}{2}.
\ee
The scaling dimension and R-charge are determined by these quantum numbers,
\be\label{sc}
h=\frac{(m+n_f)^2-j(j+1)}{k}+\frac{n^2_f}{2},~~~~~~R=\frac{2m+(k+2)n_f}{k},
\ee
with similar equations for the right movers.

The zero-mode of the dilaton, (\ref{ze}), corresponds to
\be\label{dilaton}
D: \quad (n_f=\bar{n}_f=1, m=\bar{m}=-1, j=0) .
\ee
It has $R=1$ and hence is chiral, $h=\frac12 R$.
There is another zero mode, which behaves asymptotically
as the $N=2$ supersymmetric version of Sine-Liouvile -- an `$N=2$ Liouville' mode,
with the following quantum numbers,
\be\label{tachyon}
T:\quad  (n_f=\bar{n}_f=0, m=\bar{m}=k/2, j=k/2-1) ,
\ee
that also gives $R=1$ and $h=1/2$.
This mode is the $SL(2)/U(1)$ analog of the zero mode discussed in the previous section (\ref{ta}). Indeed, it has $p=0, w=1$; it is a `winding one tachyon' mode on the cigar.
The difference is that here it appears in a setup with an exact CFT description,
which allows us to study its properties more closely.
This will be done in the next section.

The rest of the modes in the 2D harmonic oscillator of the
$w=1$ sector~\footnote{The same goes for the $w=-1$ sector.}
of (\ref{r}) are obtained by turning on the momentum $p$, which at the tip becomes the angular momentum $L$ of (\ref{o}),
\be
m=\frac12 (k+p),~~~\bar{m}=\frac12 (k-p).
\ee
Using (\ref{sc}), one can verify that the scaling
dimensions of these operators agree with the 2D harmonic oscillator of section 2, (\ref{r}),
in the $k\to\infty$ limit.

Formally, there is an agreement between the scaling dimensions and R-charges also in the $|w|>1$ sectors. These operators, however, are outside the unitarity bound, for large $k$, and so they are not part of the theory.
This conclusion illustrates the limitation of the effective action of section 2 that is not sensitive to this kind of issues.
Moreover, this could be viewed as an indication  that a closer look might reveal that our conclusion about the $w=\pm 1$ sectors is also too naive. In particular, there are two reasons to suspect that in the $k\to\infty$ limit the discrete states become part of the continuous states and that we end up with the standard string theory on $R^2$:\\
1. The dimensions of the operators in the 2D harmonic oscillator exactly agree with the dimensions of the zero momentum stringy states in flat space-time. Hence,
it is rather natural to suspect that they  become part of the continuum of $R^2$ and not twist-like operators which usually have fractural dimensions.\\
2. It turns out  that the CFT does not have two exact zero modes,
as naively implied from the discussion above, but rather a single exact zero mode~\cite{fzz,Kutasov:2000jp,Kazakov:2000pm,Karczmarek:2004bw}.
That is, the $N=2$ Liouville zero mode is combined with the zero mode of the dialton to form
a unique mode.

In the next section, we take a closer look on the second issue, and in section 5 on the first issue.

\sectiono{The zero-mode }
\bigskip

The statement that there is only one zero-mode relies on the fact that the winding number is not conserved at the tip and so it is of relevance to our discussion. The argument is the following. At the SUGRA limit, the zero mode associated with $\delta \mu$ is  (\ref{ze}).
Since $w$ is not conserved at the tip, the dilaton zero-mode can mix  with  localized operators with the same scaling dimension and R-charge and different $w$. The $N=2$ Liouville operator has exactly these properties and so on general grounds we
expect~\cite{fzz,Kutasov:2000jp,Kazakov:2000pm,Karczmarek:2004bw}
\be\label{oh}
\mu\partial_{\mu} ~~\Rightarrow ~~  -\exp(2 \Phi ) + C_T T(\rho),
\ee
where $T(\rho)$ represents the dependence of the winding one tachyon operator on $\rho$, with a specific asymptotic behavior (of an $N=2$ Liouville potential)
that fixes the normalization of $T(\rho)$, and $C_T$ is a constant.

For small $k$,
it is known that  $T(\rho)$ plays a key role everywhere (in the target space). In particular, it renders $\delta\mu$  non-normalizable for $k\leq 1$
\cite{Karczmarek:2004bw} which has important applications for string theory in 2D. Our interest here, however, is in the large $k$ limit. In this case, at least naively,  $T(\rho)$ appears negligible compared to the dilaton. In particular, far from the tip of the cigar, at the semi-infinite cylinder regime, the dependence on $\rho$ is
\be\label{asy}
\exp(2 \Phi ) \to \exp\left(-\sqrt{2/k} ~\rho \right), ~~~ T(\rho)\to \exp\left(-\sqrt{k/2} ~\rho \right) ,
\ee
and so, regardless of $C_T$, the dominant mode is the dilaton.  This is expected, as it reflects the fact that far from the tip $T(\rho)$ is a small non-perturbative $\alpha'$ correction.

For us, however, the region of interest is small $\rho$. A negligible  tachyon at large $k$ and small $\rho$ (of the order of the string scale) would mean that the tachyon got eaten by the dilaton zero mode and in that sense it becomes part of the continuum in the $k\to \infty$ limit. To determine the fate of the tachyon mode, more details are needed. In particular, we  need to fix $C_T$ and $T(\rho)$.

The constant $C_T$ was calculated in \cite{Giveon:2001up}. Taking into account the map between the cigar and Wakimoto variables~\cite{Bershadsky:1991in}, the result is (some details are presented in appendix A)
\be\label{ct}
C_T\sim k .
\ee
%
This indicates that at the tip it is probably the tachyon and not the dilaton which is the dominant factor in (\ref{oh}).
To see whether or not this is the case, we have to know $T(\rho)$.

Pure worldsheet techniques can determine only the asymptotic behaviour of $T(\rho)$ (to be (\ref{asy})).
In order to find $T(\rho)$ at the tip,
we  have to rely on target space tools -- the effective action description (\ref{ac}).
Now, because of the non-trivial geometry and dilaton, the effective action takes a slightly more complicated form \cite{Kutasov:2005rr},
\be\label{effa}
S=\frac12\int_0^\infty d\rho \sqrt{k/2} \sinh(\sqrt{2/k}~\rho)  \left((\partial_\rho T)^2+  m^2(\rho) T^2 \right) ,
\ee
with
\be
m^2(\rho)=-1+ \frac{k}{2} \tanh^2(\rho/\sqrt{2k}).
\ee
The equation of motion associated with this action has the following zero-mode solution,
\be\label{h}
T(\rho)= \frac{1}{\cosh^k(\frac{\rho}{\sqrt{2 k}})}=\frac{1}{g_0^k} \exp(k \Phi) .
\ee
By construction,
for large $\rho$ this solution agrees with (\ref{asy}),
and as expected for small $\rho$ and large  $k$ it agrees with (\ref{ta}).
The factor of $g_{0}^{-k}$  is in harmony with the KPZ scaling \cite{Knizhnik:1988ak}.

The fact that for any $k$ the  zero-mode takes such a simple form, which basically extends the KPZ scaling throughout the target space, is an indication that the use of the effective action is justified despite the issues raised in section 2.2. Worldsheet $N=2$ plays an important role in suppressing  $\alpha'$ corrections to this effective action.\footnote{For example,
in the bosonic case the analog of (\ref{ac}) does not agree with the $k\to\infty$ limit. This disagreement is correlated with the fact that in the bosonic case, unlike the $N=2$ case, the $\alpha'$ corrections are more involved than a shift $k+2\to k$ \cite{Dijkgraaf:1991ba}.}

\begin{figure}
\centering
\includegraphics[width=0.8\textwidth] {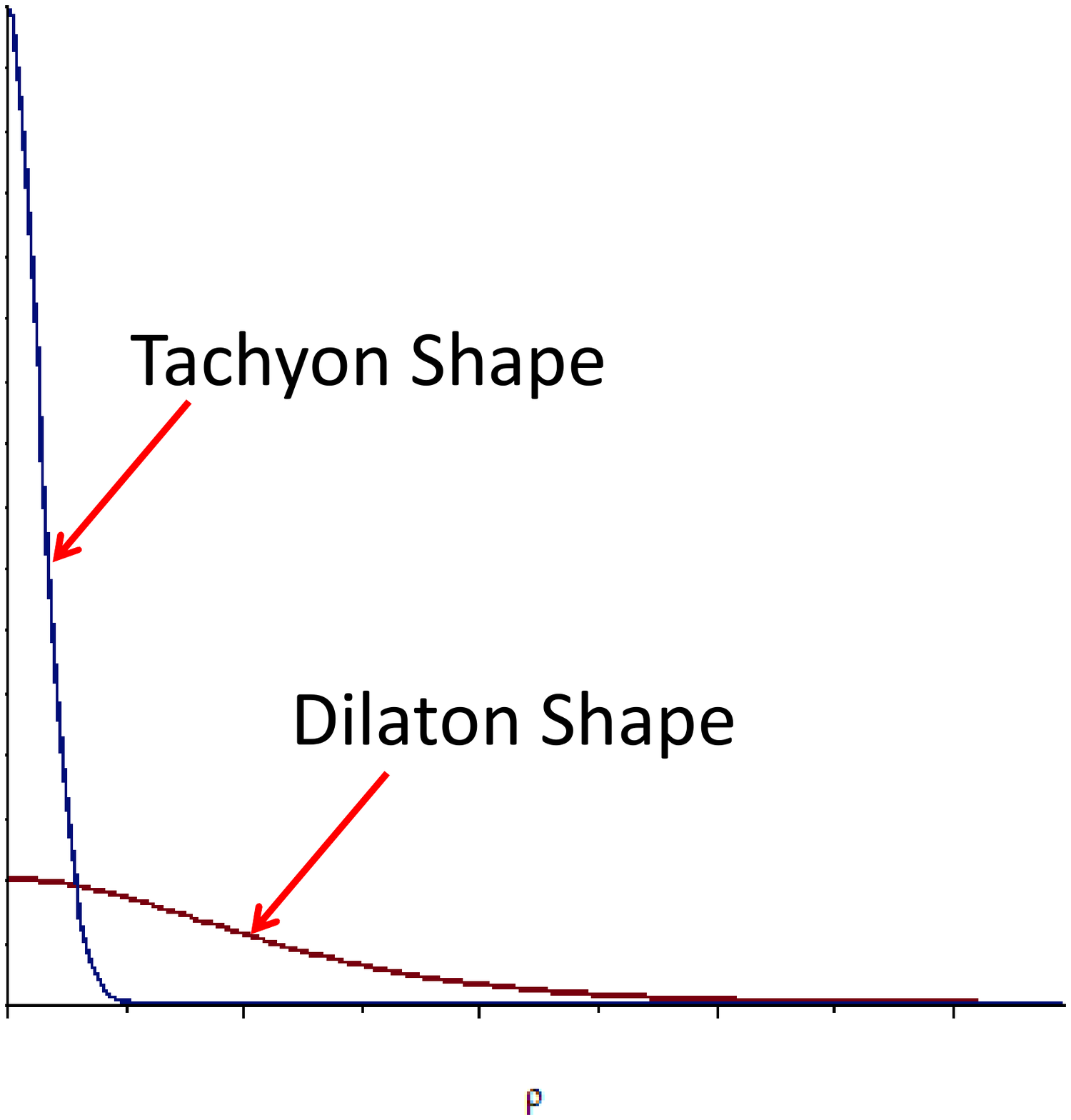}
\caption{The two components of the zero modes: the dilaton mode is the dominant mode at the cup of the cigar while the tachyon mode dominates the tip of the cigar.  \label{fzm}}
\end{figure}

Since $T(0)=1$, we conclude that at the tip the tachyon  dominates the dilaton  for large $k$
(see figure \ref{fzm}).
In particular, (\ref{ct}) implies that in the $k\to\infty$ limit, where the dilaton zero-mode becomes delta-normalizable and does not fluctuate, the tachyon still fluctuates. This is in accord with the discussion in section 2.

\sectiono{The partition function }
\bigskip

In this section, we study the partition function on $T^2$ associated with the discrete states. The motivation for doing so is two folded: First, while the arguments in the previous section are, in our opinion, quite convincing, they do rely, at least to some extent,  on the effective action (\ref{effa}) that might be misleading. This is not the case with the partition function.

The second reason is the issue raised at the end of section 3: In the limit $k\to\infty$ the quantum numbers (dimension and R-charge) associated with the discrete states
coincide with those of zero momentum modes in flat $R^2$. This seems to suggest that the discrete states might become part of the continuum of $R^2$ when  $k\to\infty$. To make this point  sharper, we recall the characters of $N=2,~c=3$ CFT's (using the notation of \cite{Klemm:2003vn}).
The characters that are standard in superstring theory on $R^2$  are the type 0 character  with $Q=0$ ($Q$ is the R-charge) and
 $h\in R ~ \backslash ~\{0\} $ ,~\footnote{The character of the identity is special, nevertheless, it
is the function $\chi^0$ with $h=0$ that appears in $R^2$.}
\be
\chi^0 = q^h
\frac{\vartheta_3 (\tau,\nu)}{\eta^3(\tau)},
\ee
and the type $I^{\pm}$  with $0<|Q|<1$ and $h=|Q|(n+\frac12)$ with $n\in  Z_{n\geq 0}$,
\be
\chi^{I^{\pm}}=q^{(n+\frac12)|Q|} z^Q \frac{\vartheta_3 (\tau,\nu)}{\eta^3(\tau)} \frac{1}{1+q^{n+\frac12} z^{sgn Q}}.
\ee
Here we focus only on the NS sector and we are using standard conventions for the $\Theta(\tau,\nu)$ functions.
The non-trivial denominator is a result of a null state \cite{Klemm:2003vn}.

The type 0 character appears in the untwisted sector while twisted sectors use  type $I^{\pm}$ characters. As discussed in section 3,
when $k\to\infty$  the discrete states have
\be
Q=\pm 1,~~~h\in  Z +\frac12.
\ee
Such quantum numbers do not appear at all in the type $I^{\pm}$ characters -- neither as primaries nor as descendants.  However, they do appear as descendants of the type 0 character. If these were the only possible characters for CFT with  $N=2$ and $c=3$, that would have been a proof that in the $k\to\infty$ limit the discrete states must become a part of the continuum of $R^2$.

This, however, is not the case. There are other, less common characters.\footnote{These characters make their appearance in the $k\to\infty$ limit
of $N=2$ minimal models, $SU(2)_k/U(1)$ \cite{Fredenhagen:2012rb,Fredenhagen:2012bw}.} The one that is relevant for us is the type $III^{\pm}$ character
\be
\chi^{III^{\pm}}=z^Q \frac{\vartheta_3 (\tau,\nu)}{\eta^3(\tau)} \left( \frac{q^h z^{sgn Q}}{1+q^h z^{sgn Q}} -\frac{q^{h+1}z^{sgn Q}}{1+q^{h+1} z^{sgn Q}} \right),
\ee
with $ Q=\pm 1,~h\in  Z +\frac12$. These are exactly the quantum numbers associated with the discrete states in the $k\to\infty$ limit.

This opens up the possibility that the discrete states do not become  part of the continuum in the $k\to\infty$ limit, but rather form a distinct sector. The partition function is a concrete way to see whether or not this is the case. If in the $k\to\infty$ limit they get absorbed into the continuum  then the modular invariant partition function associated with them should be the one of strings on $R^2$ that uses only $\chi^{0}$ and its complex conjugate. However, if they do not get absorbed into the continuum then their modular invariant partition function should take a different form.

Luckily, the finite $k$ expression for the modular invariant partition function associated with the discrete states was written recently by Sugawara \cite{Sugawara:2012ag}. Our task is to take the $k\to\infty$ limit of the expression found in \cite{Sugawara:2012ag}. Before we do that,
there are some aspects of the finite $k$ expression that we wish to review (see \cite{Eguchi:2010cb} and references therein).

The partition function associated with a sector of discrete states of closed strings can be written, schematically,  in the form
\be\label{par}
Z=\sum_{i,\bar{i}} N_{i\bar{i}} \chi_{i}(\tau) \chi_{\bar{i}}^*(\tau),
\ee
where $N_{i\bar{i}}$ are integers and $\chi_{i}$ are the relevant characters. Normally,
$\chi_{i}$ have simple modular transformations properties, and as a result it is fairly straightforward
to write down a modular invariant partition function. This is not the case with the discrete states of the  $SL(2)/U(1)$ model, where attempts to find a modular invariant partition function that takes the form (\ref{par}) failed.

A breakthrough was made by Troost \cite{Troost:2010ud},
which focused on the elliptic genus of the  $SL(2)/U(1)$ model.\footnote{To be more precise Troost \cite{Troost:2010ud} calculated the elliptic genus of the $Z_k$ orbifold of the cigar. Following \cite{Troost:2010ud} the elliptic genus of the cigar was calculated in \cite{Eguchi:2010cb,Ashok:2011cy}.} The elliptic genus is a simpler expression than the partition function,
that is holomorphic and transforms in a  covariant way under modular transformations. Troost calculated the elliptic genus  using the functional integral approach,
which by construction  guaranties a modular covariant expression. Instead of finding a holomorphic expression he encountered a holomorphic  anomaly.
That is, the modular covariant  elliptic genus contains a non-holomorphic piece.

The same goes to the partition function  \cite{Eguchi:2010cb,Sugawara:2012ag}. A modular invariant $Z_{dis}^{mod}$ can be found  using the functional integral approach and can be written in the form (\ref{par}). Only that now  $\chi_{i}$ contains a non-holomorphic piece. Schematically,
\be
\chi_i=\chi_{i}(\tau)+f_i(\tau,\bar{\tau}).
\ee
This holomorphic anomaly will play an important role below.

We start by focusing on the {\em non} modular invariant partition function associated with the discrete states. The reason is that technically it is much simpler  and much can be learned from it. The expression takes the form \cite{Eguchi:2010cb,Sugawara:2012ag}~\footnote{While our $Z$ includes
the (extended) characters of the entire $N=2$ SCFT with $c=3+{6\over k}$,
arising from the $SL(2)_k/U(1)$ piece, we
restrict here only to the contribution of free worldsheet fermions from the `internal' $SU(2)_k\times T^5$ in (\ref{slsut}) and the superconformal ghosts,
for simplicity, since the overall factor arising from the bosonic d.o.f. of internal space and the reparametrization ghosts are not important here.}
\be\label{pf}
Z_{dis}=\sum_{\sigma_{L,R}}\sum_{v=0}^{2 k-1} \sum_{a, \tilde{a}} \epsilon( \sigma_L;v,a,\tilde{a}) \epsilon( \sigma_R;v,a,\tilde{a})
\chi_{dis}^{\sigma_L} (v,a) \bar{\chi}_{dis}^{\sigma_R} (v,\tilde{a})
\left(\theta_{\sigma_L}/\eta\right)^3 \overline{\left(\theta_{\sigma_R}/\eta\right)}^3.
\ee
The left- and right-handed spin structures, $\sigma_L, \sigma_R$, run over the four sectors, $NS, \tilde{NS}, R$ and $\tilde{R}$.
The characters take the form
\be\label{cha}
\chi_{dis}^\sigma(v,a;\tau,z)=
\sum_{n\in Z}
\frac{(y q^{2kn+a})^{v/2k}}{1+(-)^{t(\sigma)}
y q^{2kn+a}} y^{4(n+a/2k)} q^{4k(n+a/2k)^2}\frac{\theta_{\sigma}(\tau,z)}{\eta^3(\tau)}~,
\ee
with $t=0~ (1)$ in the $NS~ (\tilde{NS})$ and $R~ (\tilde{R})$ sectors.
The nontrivial denominator in the character is a result of the null states \cite{Kiritsis:1986rv}.
The relationship between $v, a, \tilde{a}$ and the notation used in the previous section is
\be\label{vj}
v=2(2j+1)~,
\ee
and
\be
(a,\tilde{a})=(m+n_f-j-\frac12,\bar m+\bar n_f-j-\frac12)=(n+n_f-\frac12,\bar n+\bar n_f-\frac12)~,
\ee
with
\be
n,\bar n=1,2,...~,\qquad n_f,\bar n_f=0(\pm 1/2) ~~ \mbox{for}~~~ \sigma = NS,\tilde{NS} (R,\tilde{R})~.
\ee
Thus,
\be
a (\tilde{a}) \in \left\{\begin{array}{ll}
   Z_{2k}-\frac12 &~~ \mbox{for}~~~ \sigma_L (\sigma_R) = NS, ~\tilde{NS},\\
   Z_{2k} & ~~\mbox{for}~~~ \sigma_L (\sigma_R) = R, ~\tilde{R}, \end{array}\right.
  \ee
and
\be\label{qa}
v+2k(a+\tilde{a}) \in 2kZ.
\ee
Finally,
\be
\epsilon= \left\{\begin{array}{ll}
   1 &,~~~\sigma = NS,\\
   (-)^{1+(v+2(a+\tilde{a}))/2k}&,~~~   \sigma = \tilde{NS},\\
   -1 &,~~~\sigma = R\\
    (-)^{(v+2(a+\tilde{a}))/2k}  &,~~~ \sigma = \tilde{R}, \end{array}\right.
   \ee

The character (\ref{cha})  suggests that a continuous spectrum could emerge in the limit $k\to \infty$,
since the dimension becomes continuous. For this to happen we need to have a continuum both for $\sigma_L$ and $\sigma_R$.  The constraint (\ref{qa}) prevents this from happening.
Taking the limit $k\to \infty$ of (\ref{pf}) yields a discrete spectrum,
\ben\label{ty}
Z_{NS-NS}&=&{1\over |\eta|^{12}}\sum_{n=0}^{\infty} (2n+1) X_n^{NS} (\bar{\theta}_3^4- \bar{\theta}_4^4) +c.c.,\nonumber \\
~~~Z_{R-R}&=&{1\over |\eta|^{12}}\left(\frac14|\theta_2|^8 -2\sum_{n=0}^{\infty}n(X_n^R \bar{\theta}_2^4 +c.c.)\right),\\
 Z_{R-NS}&=& {1\over |\eta|^{12}}\sum_{n=0}^{\infty} \left(2n X_n^R (\bar{\theta}_3^4- \bar{\theta}_4^4)-(2n+1) \theta_2^4 \bar{X}_n^{NS}\right),\nonumber
\een
with
\be
X_n^{NS}=\frac{q^{n+\frac12}}{1+q^{n+\frac12}}\theta_3^4+ \frac{q^{n+\frac12}}{1-q^{n+\frac12}}\theta_4^4 ,~~~~~
X_n^R=-\frac{q^{n}}{1+q^{n}}\theta_2^4,
\ee
namely,
\be\label{ztotal}
Z_{dis}={1\over |\eta|^{12}}\left[\frac14|\theta_2|^8
+\left(\sum_{n=0}^\infty\left((2n+1)X_n^{NS}+2nX_n^R\right)\left(\bar\theta^4_3-\bar\theta^4_4-\bar\theta^4_2\right) +c.c.\right)\right]~.
\ee
Using the abstruse identity of Jacobi, we see that the spectrum admits  space-time SUSY except for the $ \frac14 |\theta_2|^8 $ term in the $R-R$ sector. This term
arises from the $\sigma_L=\sigma_R=R$, $v=a=\tilde a=0$ term in the sum (\ref{pf}),
and hence it corresponds to  $j=-1/2$ (see (\ref{vj})).
Therefore, it actually may be viewed as part of
the zero momentum modes of the continuum (see (\ref{js})).

Even if we ignore this term, we see that (\ref{ty}) is different from the zero momentum contribution to the partition function of strings on flat space,
\be\label{flat}
Z_{flat} \sim |(\theta_3^4-\theta_4^4-\theta_2^4)|^2~.
\ee
This happens despite the fact that all quantum numbers are the same as in $R^2$.
As alluded to at the beginning of this section, the partition function involves $\chi^{III^{\pm}}$, as can be seen with the help of
the relation
\be
\sum_{n=1}^\infty {q^n\over 1+q^n}=\sum_{n=0}^\infty n~ F_n~,~~~~ F_n=\frac{q^n}{1+q^n}-\frac{q^{n+1}}{1+q^{n+1}}~,
\ee
and similar relations.
The fact that (\ref{ty}) and (\ref{flat}) disagree
suggests that the modular invariant completion of (\ref{ty}) cannot be that of $R^2$. Below we verify that this is indeed the case.

For any $k$, the modular invariant partition function takes the form of (\ref{pf}), but instead of (\ref{cha}), that does not ``close" under modular transformations,
we have a `modular completed' character with a non-holomorphic correction \cite{Eguchi:2010cb,Sugawara:2012ag},
\be
\hat{\chi}_{dis}^\sigma =\chi_{dis}^\sigma -\frac12\sum_{j\in Z_4}(-)^{j(t(\sigma)-1)} R_{v+2kj,4k} (\tau) \Theta_{v+2kj+4a, 4k}(\tau,z/k) \frac{\theta_{\sigma}(\tau,z)}{\eta^3(\tau)},
\ee
with
\be
\Theta_{m,k}(\tau,z)=\sum_{n=-\infty}^{\infty} q^{k(n+m/2k)^2} y^{k(n+m/2k)}~,
\ee
and
\be
R_{m,k}=\sum_{\nu\in m+2k Z} \mbox{Erfc}\left(\sqrt{\pi\tau_2 /k} |\nu| \right) q^{-\nu^2/4k}.
\ee
The complementary error function has the standard definition
\be
\mbox{Erfc}(x)=
\frac{2}{\sqrt{\pi}}
\int_{x}^{\infty} dt\, e^{-t^2}.
\ee
Hence, we see that not only $ \hat{\chi}_{dis}^\sigma$ has a non-holomorphic piece, but that piece is not analytic in $\tau$.

The left (right) characters have a holomorphic (anti-holomorphic) piece and a non-holomorphic piece. Hence, overall we have four sectors
\be
\mbox{hol} \times \overline{\mbox{hol}}, ~~\mbox{hol} \times \overline{\mbox{non-hol}},~~\mbox{non-hol} \times \overline{\mbox{hol}},~~~ \mbox{non-hol} \times \overline{\mbox{non-hol}}.
\ee
The constraint (\ref{qa}), that prevented the first sector from having a continuous spectrum,  does not exclude the existence of a continuous spectrum in the other three sectors.
For example, the sector $\mbox{non-hol} \times \overline{\mbox{non-hol}}$ has a continuous piece,
\be\label{dc}
Z_{con}^{non-non}=\frac{k}{4|\eta|^{12}}|(\theta_3^4-\theta_4^4-\theta_2^4)|^2
\left( \int_{-\infty}^{\infty}  dp  \int_{-\infty}^{\infty} d\bar{p} (q \bar{q})^{-p \bar{p}} \mbox{Erfc}^2(\sqrt{\pi\tau_2} |p +\bar{p} |)
\right)~,
\ee
with $(p, \bar{p})= \sqrt{\frac{2}{k}} (a, \tilde{a})$.
We see that the non-analyticity survived the $k\to\infty$ limit. Momentarily,
we shall show that the other sectors do not cancel this non-analyticity. Since the partition function of flat space is analytic in $\tau$,
we conclude that the modular completion of the discrete states is not that of flat space.

Note that $Z_{con}^{non-non}$ uses only $\chi^0$ (and not $\chi^{III}$). As a result, the oscillatory part of $Z_{con}^{non-non}$ is the same as in flat space.   The momentum contribution, however, differs considerably from the standard momentum contribution on $R^2$,
\be\label{fl}
k\int dp_1 dp_2 (q\bar{q})^{\frac12 (p_1^2+p_2^2)}.
\ee
The small, $1/k^2$, curvature  cannot account for such a drastic difference that, as mentioned above, includes non-analyticity in $\tau$. $Z_{con}^{non-non}$ is modular invariant by itself, as can be seen by performing the integral over $p$ and $\bar{p}$ that yields $1/\tau_2$.

Eventually, we would like to know which string theory includes such an unusual term in its partition function.  We are far from having an answer to this question. Still we wish to make a couple of comments about $Z_{con}^{non-non}$,
that might prove useful in finding the string theory in question.
First, the asymptotic expansion of the complementary error function,
$$
\mbox{Erfc}(x) \to e^{-x^2}/x\sqrt{\pi}~,
$$
implies that, up to logarithmic corrections, the momentum piece of  $Z_{con}^{non-non}$ agrees with (\ref{fl}) in the point particle limit, $\tau_2\to\infty$. Second, it is often the case that a non-analytic expression can be written as an integral over an analytic function. In our case, we have
\be
\mbox{Erfc}(\sqrt{\pi\tau_2} p)=\frac{p}{\pi}
\int_{-\infty}^{\infty}
\frac{\exp\left(-\pi\tau_2(q^2+p^2)\right)}{q^2+p^2}dq.
\ee
Since now $q$ and $p$ appear as ordinary momenta in the exponent,
this could be viewed as an indication that the string theory we are after lives in higher dimensions. We will resist our temptation to speculate about possible implications of this observation.

A continuous spectrum appears also in the sectors $\mbox{hol} \times \overline{\mbox{non-hol}},~~\mbox{non-hol} \times \overline{\mbox{hol}}$. This can happen only when the denominator of the holomorphic (or anti-holomorphic) part is negligible. This restriction implies that a continuum emerges in this sectors only when $p$ and $\bar{p}$ have opposite signs. This is one reason why the continuum from these sectors cannot cancel $Z_{con}^{non-non}$. Another is that this continuum scales like $\mbox{Erfc}(\sqrt{\pi\tau_2} |p +\bar{p} |)$ (and not $\mbox{Erfc}^2(\sqrt{\pi\tau_2} |p +\bar{p} |)$).

Finally, note that the sum of the continuous pieces we find are not equal to the full modular completion of the discrete states partition function, $Z_{dis}^{mod}$,
in the $k\to\infty$ limit.
Rather, it is merely the piece that scales like $k$. The difference between the two
scales like $k^0$, and it includes the discrete states of (\ref{ty}) and $1/k$ corrections to the continuous pieces that results from approximating the sums by integral.

\sectiono{Discussion}

The goal of this paper was to take advantage of the exact CFT description of the $SL(2)/U(1)$ model to test the claim made in  \cite{Giveon:2012kp} that the tip of the cigar is special in string theory. We believe that the results presented here support the claims of \cite{Giveon:2012kp}. Moreover, the fact that the $SL(2)/U(1)$ model is  compatible  with the $|w|= 1$ sector of section 2 could be viewed as an indication that the String Theory at the Tip of the Cigar (STTC) is universal.

The next step is to find a useful definition of STTC. The definition used here is to  take $k\to\infty$ of the $SL(2)_{k}/U(1)$ theory
while focusing on the tip. This is a rather clumsy definition since it requires performing the calculation in the $SL(2)_{k}/U(1)$ model, that are typically quite tedious,
and then take $k\to\infty$. The fact that the  $k=\infty$ expressions are so much simpler than the finite $k$ expressions supports the expectation that there is a simpler definition of STTC.

A Wick rotation of STTC should provide us with a microscopical description of the horizon of  large  black holes in string theory and
could address the question whose urgency was increased by the work of AMPS \cite{Almheiri:2012rt}:\footnote{Less convincing reasons to wonder about this question include \cite{Itzhaki:1996jt, Braunstein:2009my}.}
{\it What is the (fire)wall made of?}

\begin{figure}
\centering
\includegraphics[width=0.7\textwidth] {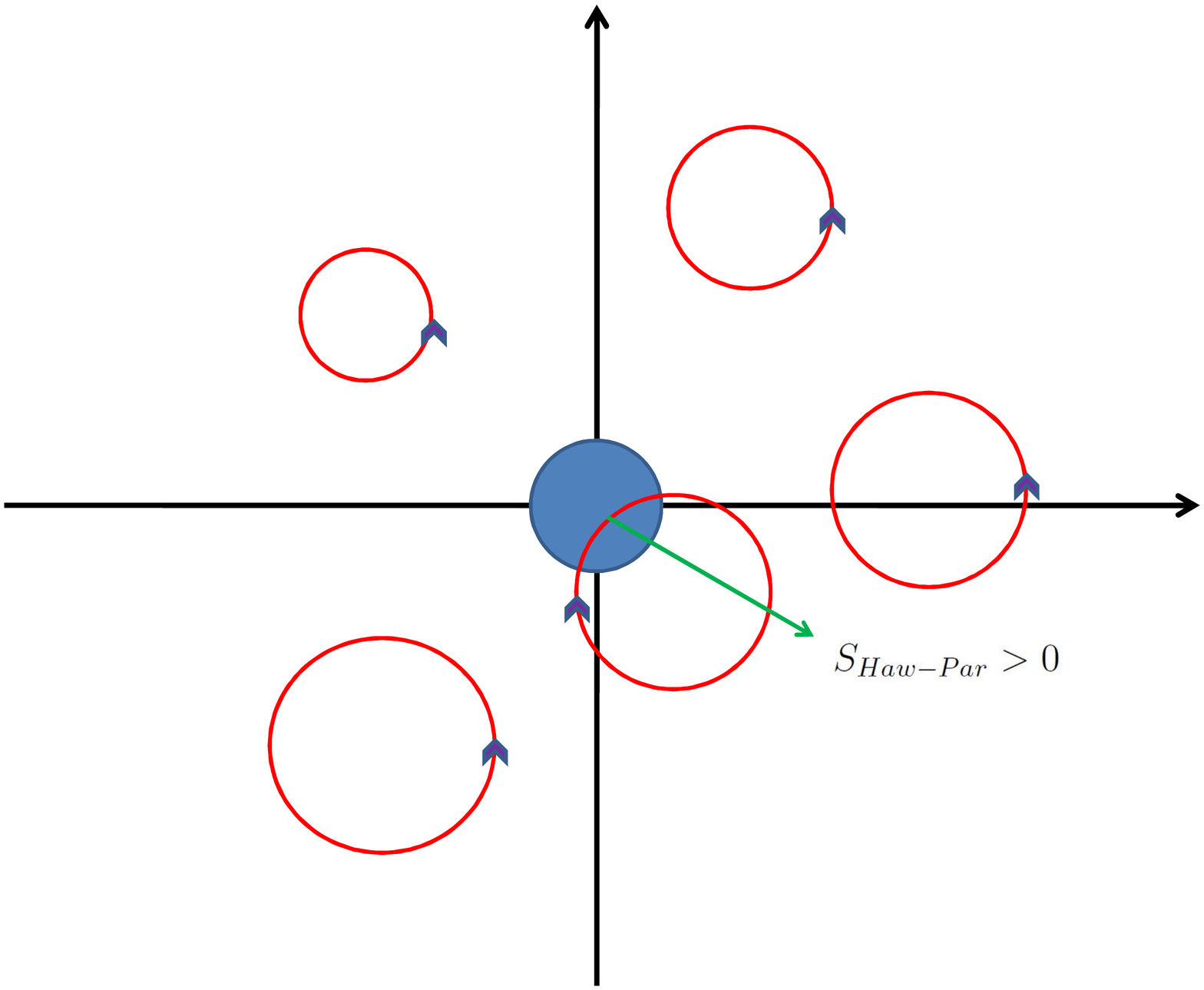}
\caption{Bubbles that do not cross the origin do not correspond to a Hawking pair and do not interact with the stringy modes at the origin. Bubbles associated with Hawking particle and its partner do cross the horizon. Their  interaction   with the stringy modes at the origin is the cause of (\ref{hp}).     \label{fbubbles}}
\end{figure}

In light of the surprising properties of STTC that we encountered so far,
we rather not speculate too much about the nature of the black hole horizon in string theory. We do wish, however,  to comment on two issues.\\
{\it 1. Firewall vs. Hawking's calculation}\\
The conclusion that in string theory there are degrees of freedom that are confined to the tip goes well with the  conclusion of \cite{Mathur:2009hf,Almheiri:2012rt} that
\be\label{hp}
S_{{\rm Hawking\, particle} - {\rm Partner}}>0.
\ee
The argument is the following \cite{Giveon:2012kp} (see figure \ref{fbubbles}).
In Euclidean space, the statement that a pair is created in a pure state is the fact that this process is described by a bubble diagram that closes without external interactions. This is what happens in ordinary $R^2$ which is the reason why an Unruh particle and its partner form a pure state and there is no firewall at the Rindler horizon. In our case, we have $R^2$ but with large fluctuations of the tachyonic field at the origin. Most bubbles  are away from the origin and will close without   interacting with the localized tachyonic modes. These bubbles, however, are not the analog of a Hawking particle (that makes it all the way to infinity) and its partner (that falls towards the singularity).
The analog of such a pair is a bubble that almost crosses the origin. Such a bubble interacts with the localized modes.
Therefore, the Hawking particle associated with such a bubble and its partner do not form a pure state in agreement with  (\ref{hp}).

It is often argued that (\ref{hp}) necessarily modifies the calculation of the black hole radiation to the extent
that Hawking's original calculation \cite{Hawking:1974sw} is no longer a good approximation. It is interesting to see what can be said about this in the Euclidean setup considered here.
In Euclidean signature, Hawking's temperature is determined by the absence of a conical singularity at the tip.
Since the action of the lowest tachyon mode vanishes,
fluctuations of this mode do not induce a conical singularity at the tip. Moreover, since the massive spectrum is supersymmetric (see (\ref{ztotal})), it cannot induce a conical singularity either. Hence, the temperature is intact. This suggests that the stringy degrees of freedom at the tip managed to induce (\ref{hp}) without modifying Hawking's temperature.\\
{\it 2. Where is the wall?}\\
We have to distinguish between two related questions. The first is where will an infalling observer realize that there is a wall? The second is a bit different: Eq. (\ref{hp}) is caused by something (the stringy modes in our case). On which side of the horizon that something is located?
Let us start with the first question. Only after crossing the horizon an infalling observer will notice that, because of (\ref{hp}), the effect of the Hawking particle is not fully cancelled by its partner. For this reason the answer to this question is just behind the horizon \cite{Almheiri:2012rt}.

On general grounds, it is clear that the answer to the second question should be just outside the event horizon
-- at the stretched horizon.\footnote{In fact, Kutasov ~\cite{Kutasov:2005rr} interpreted
`the winding tachyon condensate' region near the tip as the Euclidean version of the stringy stretched horizon of Susskind \cite{Susskind:1993ki}. This region
increases as $k$ decreases, till it is stretched all over space as $k$ approaches $1$,
when the 2D black hole (and correspondingly the zero-mode solution (\ref{h}))
turn non-normalizable~\cite{Karczmarek:2004bw}. The critical $k=1$ level is also where the black-hole/string phase transition occurs~\cite{Giveon:2005mi}. At $k=2$
(when $e^{2\Phi(\rho)}$ and $T(\rho)$ have the same asymptotic behavior, (\ref{asy})),
it was conjectured~\cite{Giveon:2006pr} that string theory on the background (\ref{slsut}) describes the type II string on a small 4D Schwarzschild black hole,
and is dual to excited fundamental strings. The behavior of the theory as $k$ decreases,
as well as the properties of the 2D black hole under T-duality~\cite{Giveon:1991sy,Dijkgraaf:1991ba},
may hint towards the nature of the wall at the horizon of the black hole.}
 The reason is that the cause of
(\ref{hp}) cannot be  behind the horizon since then it will not be able to  purify the outgoing radiation.

It is interesting to see how the Euclidean setup agrees with that conclusion. Here we considered the Wick rotation of the region outside the horizon and concluded that there are degrees of freedom at the tip. What happens if we do the same with the region between the horizon and the singularity? For Schwarzschild black holes we do not know how to address this question. However,  for near extremal $k$ NS5-branes we do. With a different motivation in mind this was already done in \cite{Fredenhagen:2012rb,Fredenhagen:2012bw}. In this case, the region between the horizon and the singularity is described by an $N=2$ minimal model with $c=3-6/k$, an $SU(2)_k/U(1)$ SCFT (see e.g. the review~\cite{Giveon:1994fu}, from where this can be deduced).
Taking $k\to\infty$ gives $c=3$, but not necessarily $R^2$. There are different ways to take the limit \cite{Fredenhagen:2012rb,Fredenhagen:2012bw}. One way is to take  $k\to\infty$ while focusing on the singularity.
This gives a continuous orbifold of $R^2$ by rotation, $C/U(1)$; its Wick rotation
-- an $R^{1,1}/Boost$ orbifold -- thus describes
the singularity of a large black hole in string theory.\footnote{With a different, though   related, motivation in mind the $R^{1,1}/Boost$ orbifold was studied extensively (see e.g. \cite{Liu:2002ft,Liu:2002kb,Horowitz:2002mw,Berkooz:2002je}).}
The other, that is relevant to us, is to take  $k\to\infty$ while focusing on the horizon. Here one  gets $R^2$ with no additional excitations. This means that the non-trivial stringy degrees of freedom live outside the horizon and not behind the horizon.

An infalling matter (or observer) that interacts directly with the cause of (\ref{hp}) will experience the firewall at the stretched horizon and not just behind the horizon.

\vspace{10mm}

\nl{\bf Acknowledgments}\\
We thank  D. Kutasov and J. Maldacena for discussions.
This work is supported in part by the I-CORE Program of the Planning and Budgeting Committee
and the Israel Science Foundation (Center No. 1937/12). The work of AG is supported in part
by the BSF -- American-Israel Bi-National Science Foundation,
and by a center of excellence supported by the Israel Science Foundation
(grant number 1665/10).

\appendix

\sectiono{$C_T$}
\bigskip

%

In this appendix, we present some comments concerning the calculation of $C_T$. The $SL(2)/U(1)$ SCFT
is described asymptotically by the theory on the cylinder
$(\rho,x_0)$ with a linear dilaton in the radial direction, $\Phi=-{1\over\sqrt{2k}}\rho$,
and interaction~\cite{Aharony:2004xn}
\be\label{lll}
{\cal L}_{\rm int}=-\lambda_d\partial x_0\bar\partial x_0 e^{-\sqrt{2/k}\,\rho}
+\lambda_{sl}\left(\psi\bar\psi e^{-\sqrt{k/2}(\rho+i\tilde x_0)}+c.c.\right)~.
\ee
Here,
$\tilde x_0=x_{0L}-x_{0R}$ is the T-dual of $x_0$. The comparison to the discussion in~\cite{Aharony:2004xn},
in particular to eqs. (2.47) and (3.9) there, is done by identifying $\phi=\rho$, $Y=x_0$ and $Q=\sqrt{2/k}$.
Note that we present here the $N=2$ Liouville case (compare to (6.6) in~\cite{Giveon:2001up} with $\psi=\psi^\rho+i\psi^{x_0}$).

The second term is the asymptotic behavior of the winding tachyon $T(\rho)$, (\ref{asy}).
The first term is the effective interaction corresponding to the metric expansion
on the cigar, (\ref{cigar}), around its large $\rho$ limit.
This interaction can be related to the Wakimoto description of $SL(2,R)$ CFT in terms of free fields. Explicitly,
$\lambda_d$ in (\ref{lll}) is related to $\lambda$ in (3.1) of~\cite{Giveon:2001up}
by considering the map between the Wakimoto variables and the cigar variables,
as can be read from eqs. (18),(19) in~\cite{Bershadsky:1991in}. One finds
$$\lambda_d\simeq k\lambda,$$
which is compatible with the expectations that
$$\lambda\simeq 1/k,\qquad \lambda_d\simeq 1.$$
The expectation that $\lambda\sim 1/k$, at large $k$,
is deduced by requiring that the Wakimoto description in eq. (3.1) in~\cite{Giveon:2001up}
gives $AdS_3$ at level $k$ (eq. (2.1) in~\cite{Giveon:2001up})
upon integrating out the Wakimoto fields $(\beta,\bar\beta)$.
The expectation that $\lambda_d$ is order one, at large $k$,
is obtained by a large $\rho$ expansion of the cigar metric (\ref{cigar}) and comparing to (\ref{lll}).

The definition of $C_T$ in (\ref{oh}) means that
\be\label{ctll}
C_T=\lambda_{sl}/\lambda_d~.
\ee
Thus, to establish the result $C_T\sim k$,
at large $k$, as discussed in section 4,
we can use the results of~\cite{Giveon:2001up}, where the following was argued.
To calculate correlation functions perturbatively in the $SL(2)/U(1)$ CFT,
one should consider the two operators in the action (\ref{lll}).
Assuming (based on experience with Liouville)
that the physics of degenerate operators is dominated in the asymptotic regime $\rho\to\infty$,
one finds, by using different degenerate operators to establish recursion relations for the two-point-functions,
that consistency leads to
\be\label{lsl}
\lambda_{sl}=-k/\pi~,
\ee
at $k\to\infty$, and
\be\label{lld}
\lambda=1/\pi k \quad\Rightarrow\quad\lambda_d\sim 1,
\ee
in agreement with $C_T\sim k$.

\end{document}